
\input harvmac
\sequentialequations

\def\BZ{{\bf Z}}
\def\BP{{\bf P}}
\def\kper{$\kappa$--period}
\def\r{\hbox{$\rho$}}
\def\p{\hbox{$\pi$}}
\def\P#1{\hbox{${\bf P}^{#1}$}}
\def\G{\hbox{$\Gamma$}}
\def\T{\hbox{${\cal T}$}}
\def\C{\hbox{${\cal C}$}}
\def\D{\hbox{$\Delta$}}
\def\O{\hbox{${\cal O}$}}
\def\CC{\hbox{$\C\times_{{\bf P}^1}\C$}}
\Title{\vbox{\hbox{HUTP-93/A008,} }
       \vbox{\hbox{RIMS-915}}      }
{\vbox{\centerline{Holomorphic Anomalies in Topological Field Theories}}}
\bigskip
\bigskip

\centerline{M. Bershadsky, S. Cecotti,\foot{On leave from
SISSA--ISAS and INFN sez. di Trieste, Trieste, Italy.}
H. Ooguri\foot{On leave from RIMS, Kyoto University, Kyoto, Japan.} and C.
Vafa}
\medskip\centerline{Lyman Laboratory of Physics}
\centerline{Harvard University}
\centerline{Cambridge, MA 02138, USA}

\vskip 1in
We study the stringy genus one partition function of $N=2$
SCFT's.  It is shown how to compute this using an anomaly in
decoupling of BRST trivial states from the partition function.
A particular limit of this partition function yields the
partition function of topological theory coupled to topological
gravity.  As an application we compute the number
of holomorphic elliptic curves over certain Calabi-Yau manifolds
including the quintic threefold.  This may be viewed as the first
application of mirror symmetry at the string quantum level.
\Date{2/93}

There has been much progress in understanding $N=2$ quantum field
theories in two dimensions in the last few years.  The main
reason for this progress has to do with the appreciation of the
existence of a topological subsector
of these theories, represented by chiral fields
($F$--terms in the supersymmetric terminology),
giving rise to a natural ring which control many aspects of these theories.
In the case of sigma models on K\"ahler manifolds chiral fields are
in one to one correspondence with the cohomology classes of the manifold
and give rise to a quantum cohomology ring.  In particular the chiral fields
are responsible for perturbations changing the
(complexified) K\"ahler class of the manifold.
These theories also have an `anti-topological' sector, which
is described by the CPT conjugate anti-chiral fields.  In addition
the theory admits an infinite number of non-topological deformations
($D$--terms in the supersymmetric terminology).  In the case of
sigma models these non-topological deformations can be
viewed as arbitrary variations of the metric leaving the
K\"ahler class invariant.

There are many computations in these
theories which do not depend on any of the infinite dimensional
non-topological perturbations
represented by the $D$--terms
and depend {\it only} on the topological ($t$) and anti-topological
($\bar t$)
parameters representing deformations of the action using chiral and
anti-chiral fields respectively (the $t$ parameters should be
viewed as holomorphic coordinates parametrizing $N=2$ QFT).
This gave rise to a new notion of a `generalized topological index'
as any computation which would be independent of
the non-topological deformations \ref\newind{S. Cecotti, P. Fendley, K.
Intriligator
and C. Vafa, Nucl. Phys. B386 (1992) 405.}\ref\ising{S. Cecotti and C. Vafa,
{\it Ising Model and N=2 Supersymmetric Theories}, preprints Harvard
HUTP--92/A044 and SISSA--167/92/EP (1992).}.  In this sense the chiral ring
(which depend only on $t$) and the anti-chiral ring
(which depend only on $\bar t$) are the simplest examples.  However
there are more interesting examples of such computations depending
on both $t$ and $\bar t$, such as the Zamolodchikov metric on
moduli space.
The basic tool to compute these new indices is the equations
derived in \ref\ttstar{S. Cecotti and C. Vafa, Nucl. Phys. B367 (1991) 359.}\
called $tt^*$ equations (topological
anti-topological equations) which capture the geometry
of the vacuum bundle as a function of the perturbations.  These
equations are a generalization of
the equations representing the variations of Hodge
structure for Calabi-Yau manifolds (known as special geometry) \ref\sp{
S. Ferrara and A. Strominger, in {\it String `92}, ed. R. Arnowitt et al.
(World Scientific, Singapore 1990), p.245\semi S. Cecotti, Commun. Math. Phys.
131 (1990) 517\semi A.
Strominger, Commun. Math. Phys. 133 (1990) 163\semi P. Candelas
and X.C. de la Ossa, Nucl. Phys. B355 (1991) 455\semi R. D'Auria, L. Castellani
and S. Ferrara, Class. Quant. Grav. 1 (1990) 1767.}\ to arbitrary $N=2$ QFT's.

For massive $N=2$ QFT's
the $tt^*$ equations were shown to be the same equations as
those capturing the $n$-point spin correlation of the Ising model \ising\
off criticality.  However the spin correlations, which are given
by the famous tau function, are constructed
non-trivially from solution to these equations.
Since the spin correlations are an intrinsic object in the Ising model
the question was raised as to what is the $N=2$ analog of the tau function;
in other words in purely $N=2$ terms, how do we think about the tau function.
This was found, in \ising , to be represented by a generalized
index which for an $N=2$ superconformal theory can be defined as
\eqn\defin{F_1=\int_\CF {d^2\tau \over \tau_2}{\rm Tr}(-1)^{\rm F}
{\rm F_L F_R}q^{\rm L_0}{\bar q}^{\rm {\bar L_0}}}
where the integral is over the fundamental domain $\CF$
of moduli space of torus, ${\rm F_{L,R}}$ denote the left and right
fermion numbers and the trace is over the Ramond sector
for both left-  and right-movers.  We have to delete
the contribution of the
ground states of the supersymmetric Ramond sector to $F_1$
($L_0=\bar L_0 =0$)
for the integral to converge.  But since we are interested
in variations of this object as a function of chiral and
anti-chiral perturbations,
this subtraction is irrelevant.  In other words $F_1$ is
only defined up to an addition of a constant.
Some properties of this index were discussed in \ising \foot{
Including the relation of this index with the generalization of
Ray-Singer analytic torsion to loop space.}.
It is our intention in this paper
to further uncover some interesting
properties of this index.

Let us first review the main result for this index derived in \ising\
(with a modification due to a subtle contact term that was missing in \ising ;
see Appendix A).  The $N=2$ algebra implies that $F_1$ is {\it
essentially} the sum of a holomorphic and an anti-holomorphic
function of moduli, except there is a slight anomaly
mixing the two expressed by
\eqn\fund{\partial_{\bar j}\partial_i F_1=\Tr (-1)^F C_i\bar C_j
-{1\over 12}G_{i\bar j}\Tr(-1)^{F}}
where $C_i$ (resp. $\bar C_j$) is the matrix representing the multiplication
of $\phi_i$ (resp. $\bar \phi_j$) on the Ramond ground states, and $G_{i\bar
j}$ is the Zamolodchikov metric (which appears in the above
due to the contribution of a contact term discussed in appendix A).
This equation is strong enough to yield $F_1$.  The idea is that
we can use the above equation to write
\eqn\genform{F_1={\rm log}\big[M(t,\bar t )f(t)\overline{f(t)}\big]}
where, as shown in \ising\
the term $M$ can be computed using \fund\
to be
\eqn\mixing{\log
M=\sum_{p,q} (-1)^{p-q}\, {p+q\over 2}
\Tr_{p,q}\big[ \log (g)\big]-{1\over
12}K\, \Tr (-1)^F}
where $p,q$ denote the left and right Ramond charges of the vacua
(which in the sigma model case on an $n$-fold range from $-n/2,...,n/2$), $g$
denotes the ground state inner product in the $p,q$ sector and $K$ is the
K\"ahler function for the Zamolodchikov metric ($\langle \bar 0 |0  \rangle
={\rm exp} (-K)$).   The $g$ can be computed using
the $tt^*$ equations \ttstar\ (which in the case of
sigma models on Calabi-Yau is a
generalization of special geometry \sp\ to arbitrary $n$--fold).
We are only left with holomorphic function $f(t)$ to be determined.  This
can in general be fixed using regularity of $F_1$ in the interior
of moduli space and once
we know how $F_1$ should behave at
the boundaries of the moduli space\foot{Note that
\genform\ and \mixing\ imply that $f(t)$ is a holomorphic
section of a certain line bundle on moduli space and $F_1={\rm
 log}\| f \|^2$.}.  This is indeed the
same idea which is used in the context of computing
threshold corrections for heterotic strings compactified
on Calabi-Yau manifolds \ref\thresh{L.J. Dixon, V.S. Kaplunovsky and J. Louis,
Nucl. Phys.
B355 (1991) 649\semi  S.Ferrara, C.Kounnas, D.L\"ust and F.Zwirner, Nucl. Phys.
B365 (1991) 431\semi I. Antoniadis, E. Gava and K.S. Narain, preprints
IC/92/50 and IC/92/51\semi
 J.--P. Derendinger, S. Ferrara, C. Kounnas and F. Zwirner, Nucl. Phys. B372
(1992) 145.}.

 To give an idea how
one may compute the behavior of $F_1$ at the boundaries
of moduli space,
 let us consider an $N=2$ SCFT arising from
a supersymmetric sigma model on a Calabi-Yau manifold $M$ of complex
dimension $n$, and let us take the $t$'s to correspond to complexified
choices for the K\"ahler class of the manifold.  To be precise
we mean that the
K\"ahler class of the manifold is taken to be
$$k =\sum_{i} (t_i+\bar t_i)k_i$$
where $k_i$ span a basis for $H^{1,1}(M,{\bf Z})$.
Then the large
$t,\bar t\rightarrow \infty$ of $F_1$ can be computed, using the fact
that in this limit only constant maps dominate the path integral
and the leading term comes from integrals over the zero modes
of the bosonic and fermionic terms in the action.  One has
\eqn\what{\eqalign{&\Tr\Big[(-1)^F {\rm F_L F_R} q^{L_0}\bar q^{\bar L_0}
\Big]\Big|_{\infty}
={1\over (2\pi \tau_2)^n}\int d\mu
\left(\prod_r d\psi^{\bar r}
d\psi^r d\chi^{\bar r} d\chi^r\right) \times\cr
&\hskip 2cm \times
g_{i\bar \jmath}\,\psi^i\psi^{\bar \jmath}\,
g_{k\bar l}\, \chi^k\chi^{\bar l}
\exp\big[-\tau_2 R_{i\bar\jmath k\bar l}
\psi^i\psi^{\bar\jmath}\chi^k\chi^{\bar l}\big],\cr} }
where $|_{\infty}$ means the
contribution in the limit $t,\bar
t\rightarrow \infty$ which comes
only from constant maps, $n$ is the
complex dimension of the
target space and $d\mu$ its volume form.
The simplest way to get eq.\what\ is to realize that the
limit $t,\bar t\rightarrow\infty$ is
just the classical theory, and then
use the corresponding classical ensemble
to evaluate \what\foot{In fact,
the rhs of \what\ is just the classical $1d$ computation.}.
The rhs of \what\ can be rewritten as
$${(-1)^{n-1}\over (n-1)! (2\pi)^n\, \tau_2}\int d\mu \,
g_{i\bar \jmath}\,\psi^i\psi^{\bar \jmath}\,
g_{k\bar l}\, \chi^k\chi^{\bar l}\big(R_{i\bar\jmath k\bar
l}\psi^i\psi^{\bar\jmath}\chi^k\chi^{\bar l}\big)^{n-1}.$$
Integrating away the fermions we get
$${(-1)^{n-1}\over (2\pi)^m (n-1)!\, \tau_2}\int d\mu \,\epsilon^{i_1\dots
i_n}\epsilon^{\bar\imath_1\dots \bar\imath_n}\epsilon^{j_1\dots j_n}
\epsilon^{\bar\jmath_1\dots \bar\jmath_n} g_{i_1 \bar \imath_1}
g_{j_1\bar\jmath_1} R_{i_2\bar\imath_2 j_2\bar\jmath_2}\dots R_{i_n\bar\imath_n
j_n\bar\jmath_n}.$$

Next, recall the formula for the $k$--th Chern class of a (complex) manifold
$M$
\eqn\chern{c_k(M)={(-1)^k \over (2\pi i)^k k!}\delta^{j_1\dots j_k}_{i_1\dots
i_k}
R^{i_1}_{j_1}\wedge \cdots \wedge R^{i_k}_{j_k},}
since
$$\epsilon^{i_1\dots i_n} \epsilon_{i_1 k_1\dots k_n}=\delta^{i_2\dots
i_n}_{k_2\dots k_n},$$
we have the identity
$$\eqalign{&(-1)^{n-1} k \wedge c_{n-1}=\cr
&={2(-1)^{n-1}\over (2\pi)^n(n-1)!}d\mu\, \epsilon^{i_1\dots
i_n}\epsilon^{\bar\imath_1\dots \bar\imath_n}\epsilon^{j_1\dots j_n}
\epsilon^{\bar\jmath_1\dots \bar\jmath_n} g_{i_1 \bar \imath_1}
g_{j_1\bar\jmath_1} R_{i_2\bar\imath_2 j_2\bar\jmath_2}\dots R_{i_n\bar\imath_n
j_n\bar\jmath_n},\cr}$$
where $k=2i g_{i\bar \jmath}dX^i\wedge d\bar X^j$.
Using this in \what, we get
$$\Tr\Big[(-1)^F {\rm F_LF_R} q^{L_0}\bar q^{\bar L_0}\Big]\Big|_{\rm \infty }=
{(-1)^{n-1}\over (4\pi)\, \tau_2}\int\limits_M k \wedge c_{n-1}(M).$$
Then the leading term of $F_1$ in the $t,\bar t\rightarrow \infty$ is
equal to
\eqn\large{\eqalign{\int\limits_\CF {d^2\tau \over\,\tau_2 }\Tr&\left[(-1)^F
{\rm F_LF_R} q^{L_0}\bar q^{\bar L_0}\right]\Bigg|_{t,\bar
t\rightarrow\infty}=\cr
&=(-1)^{n-1}\int\limits_M k \wedge c_{n-1} \int\limits_\CF {d^2\tau \over 4\pi
(\tau_2)^2}= {(-1)^{n-1}\over 12}\int\limits_M k \wedge c_{n-1}\cr}}

In the special case $M$ is a (smooth) CY $3$--fold one has
$$\int k\wedge c_2= {1\over 8\pi^2}\int\Vert R\Vert^2 d\mu\geq 0,$$
and so the leading term vanishes if and only if $M$ is flat.

Let us consider the example of a target space being a
one dimensional complex
torus which we denote by $T^2$.
It is well known that the K\"ahler moduli of the
torus is equivalent to its complex moduli and both
are parametrized by the upper half plane up to the action
of $PSl(2,Z)$.  Let us denote the K\"ahler class of $T^2$ by $\sigma $
and its complex structure by $\rho $.  This definition
of $\sigma $ differs from that discussed above by $t=-2\pi i \sigma$.
Then applying \fund\ to this case we see that
$$\partial_\sigma \partial_{\bar \sigma}F_1={-1\over (\sigma -\bar \sigma
)^2}$$
 Therefore
we immediately learn that
$$F_1=-\log \big(\sigma_2 |f(\sigma )|^2\big)$$
where $\sigma_2$ is the imaginary part of $\sigma$.
We are now left to compute $f(\sigma )$.  However,
we know that the moduli space of $\sigma$ can be taken
to be the standard fundamental domain on the upper half plane,
and $F_1$ should be a well defined function on this domain
(which fixes the modular weight of $f$) and that for all $\sigma
\not= \infty$ it should be finite.
We learn therefore that
\eqn\torcas{F_1=-\log \sigma_2 |\eta ^2(q)|^2}
where $q={\rm exp}(2\pi i \sigma )$ and
$\eta$ is the Dedekind eta function.  Note that the leading behavior
of $F_1$ for large $\sigma$ is also in accord with that predicted
by \large\ (with $n=1$, and $\int k =[-2\pi i\sigma +c.c.]$).

\newsec{Topological Limit}

Before turning to computations for $F_1$, we will connect our
index to the genus one partition function of
topological sigma model coupled to
topological gravity \ref\wit{E. Witten, Nucl. Phys. B340
(1990) 281.}.
To this end let us consider the limit $\bar t \rightarrow \infty$ while
fixing $t$ at a finite value.  Formally this is what
one would expect to be the relevant contribution for the topological
theory which weighs only holomorphic maps from the world sheet
to the target space.  The reason $\bar t \rightarrow \infty$
accomplishes this is that in this limit the action is infinite unless
we have a holomorphic map ( otherwise
 ${\bar t}\int k_{i\bar j}{\bar \partial} X^i \partial {\bar X^j}\rightarrow
\infty$).

Let us first examine the case when the target
space is the one dimensional complex torus. We have already
computed this in \torcas\ but it is instructive to derive
this more directly by explicitly computing
$\Tr (-1)^F {\rm F_L F_R} q^{L_0} \bar{q}^{\bar{L}_0} $. Since the non-zero
modes of
the bosons and
the fermions cancel each other, it
is expressed as a sum over instantons as
$$\eqalign{{\Tr } (-1)^F {\rm F_L F_R} q^{L_0} \bar{q}^{\bar{L}_0}
      = {t+\bar{t} \over 4 \pi \tau_2}
   \sum_{m,n,r,s}&
   {\rm exp}\Big[-{t \over 4 \tau_2 \rho_2}
               |(m+r\rho)-\bar{\tau}(n+s\rho)|^2-\cr
              & -{\bar{t} \over 4 \tau_2 \rho_2}
               |(m+r\rho)-\tau(n+s\rho)|^2 \Big] \cr} $$
where $\rho$ is the complex modulus of the target torus.
In the $\bar{t} \rightarrow \infty$ limit, this becomes
\eqn\largetbar{ \eqalign{& {\Tr } (-1)^F {\rm F_L F_R} q^{L_0}
\bar{q}^{\bar{L}_0}
      \simeq\cr & \simeq {t + \bar{t} \over 4 \pi \tau_2}
   + \sum_{M \in GL (2,{\bf Z}) }
     {\tau_2 \over |\det M|} e^{-|\det M| t}
     \delta(\tau - M(\rho)) + O(e^{-\bar{t}}) \cr} }
where $M(\rho) = (m+r \rho)/(n+s \rho)$
for $M= \left( \matrix{ n & s \cr m & r \cr} \right)$.
The first term in the right-hand side corresponds
to the zero instanton sector ($m,n,r,s=0$) while
the second term represents the sum over
holomorphic instantons. In fact, when $\tau = M(\rho)$,
there exists a holomorphic map of degree $|\det M|$
given by
$$X(z) = (n+s\rho)z.$$
By integrating \largetbar\ over the moduli space
of the worldsheet torus, we obtain\foot{
We are computing the derivative of $F_1$ with
respect to $t$ since $F_{1}$ itself contains an infinite
constant as mentioned before. In general, the infinity arises in $F_{1}$
when ${\Tr } (-1)^F {\rm F_L F_R} q^{L_0} \bar{q}^{L_0}$
does not vanish at $\tau_2 \rightarrow 0$.}
\eqn\torusinstanton{ \eqalign{ {\partial \over \partial t} F_{1}^{top}
   & \equiv \int {d^2 \tau \over \tau_2} {\partial \over \partial t}
       {\Tr } (-1)^F {\rm F_L F_R} q^{L_0} \bar{q}^{\bar{L}_0}
\Big|_{\bar t\rightarrow \infty}\cr
   & =
      {1 \over 12}
        - \sum_{M: M(\rho) \in {\cal F}}
           e^{-|\det M| t}
       . \cr} }
Here the sum in the right-hand side is
over $M \in {\sl GL}(2,{\bf Z})$ such that
$M(\rho)$ is in the fundamental domain ${\cal F}$
of the moduli space. Equivalently, one may
sum over elements of ${\sl GL}(2,{\bf Z})$
with the identification $M \sim UM$ for
$U \in {\sl SL}(2,{\bf Z})$. A theorem by
Hermite \ref\herm{M. Newman, {\it Integral Matrices}, Pure and Applied
Mathematics, vol 45, Academic Press, New York 1972.}\ states that we can always
find a canonical
representative in the equivalence class of $M$ of the form
$$  \left( \matrix{ n & 0 \cr m & \pm r \cr} \right),
               ~~~n,r \geq 1,~m=0,...,n-1$$
and that this representative is unique. Thus the
above becomes
$$ \eqalign{ {\partial \over \partial t} F_{1}^{top}
     & = {1 \over 12}
       -  2\sum_{n,r=1}^\infty \sum_{m=0}^{n-1}  e^{-nrt}  \cr
     & =
         {1 \over 12}
        -2 \sum_{n=1}^\infty { n e^{-nt} \over 1 - e^{-nt}}
     = -2 {\partial \over \partial t}
        \log \eta \big( \exp (-t) \big)
       . \cr} $$
This agrees with the expression \torcas\ we have obtained by
solving the differential equation \fund\
as $\bar t\rightarrow \infty$.
It is amusing to note that this result is along the same lines as those of
 \ref\gro{D. Gross and W. Taylor, {\it Two Dimensional QCD is a String Theory},
hep-th@xxx.lanl.gov 9301068 .}\
in connecting the large $N$ expansion
of pure 2d QCD viewed as a string theory.  Indeed the
computation above is essentially the same as that in \gro\
but now with a reinterpretation in terms of a topological matter
(represented by a target torus) coupled with topological gravity.
It is tempting to conjecture that the two are indeed the same
for arbitrary target space\foot{This observation and the above
counting of holomorphic maps from the torus to the torus was also
made independently in \ref\dij{R. Dijkgraaf, R. Rudd,
in preparation.}.}.

In the case of the torus, there is in fact
only one primitive instanton $X(z)=z$ which
appears at $\tau=\rho$, and all other
instantons at $\tau= M(\rho)$ ($M \neq {\bf 1}$)
may be regarded as its multiple cover.
Since the target space torus can be viewed
as an infinite plane divided by translations generated by
$1$ and $\rho$, an image
of the worldsheet should be on a fundamantal domain of
a lattice defined by $n$ and $m \pm r \rho$ for some
$n,r \geq 1, m=0,...,n-1$ (upto the modular
transformation on the target). In this case,
the modulus $\tau$ of the worldsheet
is given by $\tau = (m+r\rho)/n$, and the holomorphic
map is $X(z) = nz$. This corresponds
to the matrix $M= \left( \matrix{ n & 0 \cr m & \pm r \cr} \right)$
and it explains why a matrix of this form
gives a canonical representative of the equivalence
relation.
As one can see in the expansion \torusinstanton,
all these instantons
are counted with multiplicity one in $\partial_t F_{1}^{top}$
independently of their degrees.  Note also the factor of 2
in front of ${\rm log}(\eta )$ simply expresses the fact
that for each holomorphic map there is another one
obtained from it by sending $z\rightarrow -z$.
 So the moral of the story is that in the case
of target space a torus there is only {\it one} primitive
elliptic curve.  All the other ones are multiple covers of it
which just recapture the geometry of the $\eta$ function.

Let us consider the more general case when the target
space is an $n$-dimensional Ricci-flat K\"ahler manifold $M$.
To start with, let us suppose there exists a holomorphic map $X_0^i(z)$ from
the worldsheet
to the target at a special value of the worldsheet
modulus $\tau=\tau_0$.
In order to evaluate its contribution to the path integral
in the limit $\bar{t}_a \rightarrow \infty$, we
expand the non-linear $\sigma$ model action $S$
upto the second order in $\tau-\tau_0$ and $x=X-X_0$ as
$$
 \eqalign{ S & = \sum_a (t_a +\bar{t}_a
       \left( {|\tau-\tau_0| \over 2 \tau_2} \right)^2
       ) s_{a}(X_0) +\cr
        & + \sum_a \bar{t}_a \left(
  \int k_{i\bar{j}}^{(a)} \partial x^{\bar{j}} \bar{\partial} x^i
     + {i \over 2} \psi_L^i \bar{\partial} (k_{i\bar{j}}^{(a)}
         \psi_L^{\bar{j}})
     + {i \over 2} \psi_R^{\bar{j}} \bar{\partial}
         (k_{i\bar{j}}^{(a)} \psi_R^i) \right) d^2z
     +O({1 \over \bar{t}}) \cr} $$
where $s_{a}(X_0)= \int k_{i\bar{j}}^{(a)}
\partial X^i_0 \bar{\partial} X^{\bar{j}}_0$ are
integers given by the homology class of the image.
The action for $x$ and $\psi_L, \psi_R$ becomes free in this
limit and is invariant under the reduced BRST transformation
$(\delta x^i, \delta x^{\bar{i}}) = (i\epsilon \psi_L^i,
i \bar{\epsilon} \psi_R^{\bar{i}})$,
$(\delta \psi_R^{i}, \delta \psi_L^{\bar{i}})
 = (2 \epsilon \partial x^i , 2 \bar{\epsilon}
    \bar{\partial} x^{\bar{i}})$.
This guarantees that the non-zero modes of
$x$ and $\psi$ cancel as in the torus
example.

When the holomorphic map $X_0$ corresponds to an isolated elliptic curve in
$M$, there is one fermion zero mode for each of
$\psi_L$, $\bar{\psi}_L$, $\psi_R$ and $\bar{\psi}_R$,
and they are soaked up by the insertions of ${\rm F_L}$ and ${\rm F_R}$.
If the map has moduli, there are additional fermion
zero modes corresponding to deformations of
the curve. Therefore such $X_0$ does not contribute
to ${\sl tr}(-1)^F {\rm F_L F_R} q^{L_0} \bar{q}^{\bar{L}_0}$
to the leading order in $\bar{t}$. In the following,
we only consider the situation when the curve is isolated.

Let us evaluate the zero mode integral.
If the holomorphic map $X_0(z)$ is primitive,
we can choose local coordinates near the image
of $X_0(z)$ in the target  such that
the coordinate along the image coincide with that of
the worldsheet. With this choice of coordinates,
the zero mode integral becomes essentially two-dimensional
since there is no fermion zero mode in the directions
normal to the image of $X_0(z)$.
Therefore the zero mode integral with ${\rm F_L}$ and ${\rm F_R}$
insertions gives the area of the image of $X_0(z)$ in the target space
divided by $4 \pi \tau_2$. If the holomorphic map $X_0$
is primitive, the map is injective and the area is given by
$\int g_{i\bar{j}} \partial X_0^i \bar{\partial} \bar{X}_0^{\bar{j}}
d^2 z = \sum_a (t_a + \bar{t}_a)s_{a}(X_0)$. On the other
hand, if the map is an $N$-fold cover of a primitive map,
it may be viewed as the composition of the primitive instanton
with an $N$-fold covering of the torus by torus, discussed before.
Therefore, as we saw before,
we must divide this by $N$ (because of the $U(1)$ isometry
of the torus).

Thus we find that, in
the limit of $\bar{t}_a \rightarrow \infty$,
the contribution of the instanton $X_0(z)$
to ${\sl tr}(-1)^F {\rm F_L F_R} q^{L_0} \bar{q}^{\bar{L}_0}$
is given by
$$ \eqalign{& \left(
 {\sum_a (t_a+ \bar{t}_a) s_a(X_0) \over 4 \pi \tau_2 N} \right)
       e^{-\sum_a t_a s_a(X_0)}
    \left( {4 \pi (\tau_2)^2 \over \sum_a \bar{t}_a s_a(X_0) } \right)
     \delta(\tau-\tau_0) \cr &
    \simeq \left( {\tau_2 \over N} e^{-\sum_a t_a s_a(X_0)} \right)
          \delta(\tau-\tau_0),\cr} $$
where the $\delta $--function (with its prefactor) emerges
from taking the $\bar t \rightarrow \infty $ of ${\rm exp}(-S)$.
 Then the contribution to $F_{1}^{top}$ obtained by
integrating the above integrand over the moduli space of tori
with measure $d^2\tau /\tau_2$ is given by
\eqn\elliptic{\eqalign{ \sum_{elliptic\atop instantons}
      {1 \over N} e^{-\sum t_a s_a}
    & = 2 \sum_s d_s \sum_{n,r=1}^\infty
         \sum_{m=0}^{n-1} {1 \over nr} e^{- nr \sum_a
            t_a s_{a} } \cr
    & = -2 \sum_s d_s\log \prod_{n=1}^\infty
               (1-e^{- n \sum_a t_a s_{a}}) \cr} }
where the integers $s_a$ label
the homology class of the elliptic curve and $d_s$ denotes
the number of curves in the class.

However \elliptic\ is not the full story. Other configurations contribute to
$F_{1}^{top}$. For instance, any meromorphic function on the torus will give us
a map from the world--sheet to a sphere which, composed with a genus zero
instanton, gives an instanton whose image is a rational curve in $M$. Thus
$F_{1}^{top}$ should also get contributions from the rational curves of $M$.
In fact the real story is even subtler. We also get a contribution from the
rational curves in degree $1$, although there is no degree $1$ meromorphic
function.  The point is that in order to reduce the path integral to a sum over
the minima
of $S$ we must compactify the integration space\foot{A convenient
compactification of `the space of all maps' was introduced by Gromov
\ref\gromov{M. Gromov, Invent. Math. 82 (1985) 307.}.} (or, equivalently, take
into account the saddle points at infinity).

Let us make the simplifying assumption
that we are dealing with a threefold, in which case the rational
curves are (generically) rigid (to avoid integration
over the moduli of rational curves).
Suppose we want to construct a
degree one instanton from the torus into a sphere in $M$. We can think of
constructing an approximate one by `gluing' at one point $z_0$ on the torus a
standard instanton for the plane having a size much smaller than
the periods of the torus. Of course, this is not an exact solution of the
equation of motion. But it becomes so in the limit of vanishing size, where we
get a delta--function instanton. Alternatively, while taking the size to zero
we can make a compensating conformal transformation of a neighborhood of $z_0$,
so that the instanton looks of finite size. In this case the world sheet will
look as a sphere attached by the point $z_0$ to the torus, the sphere being
mapped into the given rational curve of $M$ while the torus gets mapped into
$X_0(z_0)$. This is the `bubbling' phenomenon discussed in
refs.\gromov\ref\nnnn{J.G. Wolfson, J. Diff. Geom. 28 (1988) 383.}. This
phenomenon is crucial  to get the correct answer for the instanton correction
(even in genus zero). For a given rational curve, the leading contribution to
the topological one--point function comes from the `single bubble'
configuration. It is not too hard to guess what its contribution should be.  A
delta--function instanton introduces a second puncture on the torus and hence
one has to integrate over the corresponding moduli space, getting a factor
$1/12$ [$=\chi({\cal M}_{1,2})$]. The integral over the sphere then gives a
factor of the degree of the corresponding rational curve.
Finally, given that $F_1^{top}$ counts each elliptic curve
twice (by $\BZ_2$--symmetry) it is also natural to expect
a similar factor here.
The contribution to
$\partial_a F_{1}^{top}$ from these rational instantons in
the threefold case is then expected to be
(by a natural extrapolation of the genus zero result for
the contribution of multiple cover of rational curves)
\eqn\rational{-{2\over 12} \sum_s n_s s_a {e^{-\sum t_{a'} s_{a'}}\over
1-e^{-\sum
t_{a'} s_{a'}}},}
where $n_s$ denotes
the multiplicity of primitive holomorphic maps from sphere
to $M$ in the given homology class.
It can be shown (see appendix B by S. Katz)
that \rational\ is the natural formula for the rational contribution to
$\partial_a F_{1}^{top}$ from the viewpoint of  Algebraic Geometry in the
spirit of ref.\ref\ratcurves{P.S. Aspinwall and D.R. Morrison, {\it Topological
Field Theory and Rational Curves}, Oxford preprint OUTP--91--32p,
DUK--M--91--12 (1991).}.

Combining \elliptic\ with \rational\ and the contribution from the zero
instanton
sector which was evaluated before, we obtain
$$ \eqalign{
 {\partial \over \partial t_a}& F_1^{top}
     = {(-1)^{n-1} \over 12} \int k_{a} \wedge c_{n-1}-\cr
     & - 2 \sum_s d_s\sum_{n=1}^\infty
           { n s_a
           e^{-n\sum t_{a'} s_{a'}} \over 1 -
      e^{-n \sum t_{a'} s_{a'} }}
-{2\over 12} \sum_s n_s s_a {e^{-\sum t_{a'} s_{a'}}\over 1-e^{-\sum t_{a'}
s_{a'}}}\cr}$$
\eqn\betfor{=-2 {\partial \over \partial t_a}\Big[\sum_s [d_s
\log {\eta (q^s)}+{1\over 12}n_s\log (1-q^s)]\Big] +{\rm const.}}
where $q^s={\rm exp}(-t_a s_a)$ and $d_s,n_s$ are the number
of primitive holomorphic maps from torus and sphere to $M$, respectively.
The torus
example is the case where $d_s=n_s=0$ for all $s$ except $d_1=1$.

We have thus seen that in the limit as $\bar t\rightarrow \infty$,
$\partial _iF_1$ essentially counts the number of elliptic curves in the target
space such that one point of the torus is mapped to the cycle
dual to the K\"ahler class $k_i$.  This is precisely
the definition of topological gravity coupled to topological
sigma model \wit .  In other words in the limits as
$\bar t\rightarrow \infty$ we can view $F_1$ as computing
the partition function of topological gravity coupled to topological
sigma model.  However, we have learned more than that here:
We have learned that the partition function of topological gravity
coupled to topological matter has a deformation in terms of anti-chiral
fields parametrized by $\bar t$.  This is a surprise in that
$\bar t$ perturbations are BRST trivial in the topological
set up.  In the topological $\sigma$-model without
coupling to the gravity, it is known
that the instanton approximation is exact.  What we
are learning here is that the story changes dramatically
when we couple to topological gravity.  In this case
perturbing the topological
theory coupled to gravity, by BRST trivial operators nevertheless
modifies the partition function.  In other words we have discovered
that {\it there is a topological anomaly in the sense that} BRST
{\it trivial states fail to decouple and
there is a finite boundary contribution}.  Indeed the
equation \fund\ could be viewed as precisely expressing this anomaly!
It is simply expressing the fact that in the definition of $F_1$
we have an unavoidable mixing between $t$ and $\bar t$.
Luckily, as is the situation with anomalies, it is precisely
for this reason that we are able to compute $F_1$ at all!

Having set  up all this machinery we now apply it to some
examples for low dimensions.

\newsec{Low Dimensional Examples and Quantum Mirror Symmetry}
After the one dimensional case, we come to the case of K3.
This is a simple case as the right hand side of \fund\
vanishes as can be seen from \mixing.  For $K3$ the leading term
\large\ also vanishes identically since $c_1(K3)=0$.
 Therefore
$F_1=0$ (due to the quasi-compactness of moduli space).  This is consistent
with the fact that there
are no holomorphic elliptic curves\foot{The contribution from rational curves
vanishes, as it does in genus zero.} on a {\it generic} $K3$. Indeed, there are
(non--algebraic) $K3$'s having no non--constant meromorphic functions
\ref\koda{K. Kodaira, Amer. J. Math. 87 (1964) 751.}. One shows (see
\ref\kodaira{K. Kodaira, Ann. Math. 74 (1961) 591.} Th.5.1) that on such a K3
there is only a finite number of irreducible curves $C_k$. Moreover the sum of
their genera is given by the formula
$$\sum_k g(C_k)\leq q-p_g+1\equiv 0,$$
and then $g(C_k)=0$ for all $k$'s. So the first non-trivial case is
a threefold CY.

Let us first write the special form \genform\ takes
in the case of the threefold.  Let us focus on the K\"ahler
moduli (which by mirror symmetry is general enough).
Let $K$ denote the K\"ahler
function for the Zamolodchikov metric $G_{i\bar j}=
g_{i\bar j}/ g_{0\bar 0}=\partial_i
{\bar \partial_j}K$ on the K\"ahler moduli.  Then from \mixing\ we get
\eqn\spca{F_1=\log\Big[ {\rm exp}\big[(3+h_{1,1}-{\chi \over 12})K\big]\
\det G_{i\bar j}^{-1}|f|^2\Big] }
where $\chi$ is the Euler characteristic of the CY and $h_{1,1}$
is the dimensions of $H^{1,1}(M,{\bf Z})$
and $f$ is some holomorphic function to be determined
by imposing appropriate boundary conditions at $F_1$ and using
the fact that it should be finite in the interior of the
moduli space.

The simplest non-trivial examples of threefolds
are the toroidal orbifolds.  The story is essentially
identical to what is known for the threshold corrections
\thresh .
In this case it is easy
to see that as long as we are interested in the K\"ahler
moduli of the underlying torus, $F_1$ will be independent
of it in case no group element leaves a one dimensional
complex torus fixed (the right hand side of \fund\ vanishes
for such cases) or proportional to the
result for one dimensional torus, in case there is a
fixed torus ($F_L,F_R$ absorb the fermion zero modes of the
fixed torus) and we end up getting (in the simplest orbifolds)
$$F_1=\sum {c_i}\log\big(\tau^i_2 \eta^2(\tau^i)\bar \eta^2(\bar \tau^i)\big)$$
where $c_i$ are some easily computable numbers, and $\tau^i$ refer
to the K\"ahler moduli of the three tori of the orbifolds.
Of course if we are interested in the dependence on blow
up modes life is more complicated.  Therefore if we
ignore the blow up modes, the situation is hardly more
interesting than the one dimensional case discussed before.

For a more interesting example let us consider the case
of the quintic threefold.  Our considerations here
in fixing the holomorphic piece of $F_1$ are very similar
to those worked out in \ref\dix{L. Dixon, Talk Presented in Mirror
Symmetry Conference at MSRI, 1991, unpublished.}\ in the context of
threshold corrections.  Using the general form \spca\
and noting that $\chi =-200$ and $h_{1,1}=1$ we have
$$F_1=\log \left\{G_{\psi \bar \psi}^{-1}\, {\rm exp}\Big[{62\over 3}K\Big]\
|f(\psi )|^2 \right\}$$
  In this case the explicit form of $K$
is worked out in \ref\cand{P. Candelas, X.C. de la Ossa, P.S. Green and L.
Parkes, Nucl. Phys. B359 (1991) 21.}\
using the mirror symmetry discovered for the quintic
in \ref\greple{B.R. Greene and M.R. Plesser, Nucl. Phys. B338 (1990) 15.}\
and we are using their results as well as their notation.
Here we are applying this symmetry at the string one-loop level,
which is thus at the quantum level\foot{
In the sense we are using the word {\it quantum} here, the quantum
cohomology ring should be viewed as the {\it classical} string
computation, i.e., on the sphere.}.
The modulus parameter $\psi$ describes a degenerate
CY only at $\psi=1,\infty$.  So regularity implies
that $F_1$ should be finite everywhere except possibly
at these two points.  Moreover, multiplication of $\psi$
by a fifth root of unity is a modular transformation
and should leave $F_1$ invariant.  Finally by
general arguments we know how $F_1$ should behave
in the large volume limit, which for us is the
limit $\psi \rightarrow \infty $ at which limit
$\psi^5 \sim {\rm exp}(t)$.  Using \large\ and
the simple computation which shows
$$\int k\wedge c_2 =50$$
we learn that the large $\psi$ limit of $F_1$ should
be given by
\eqn\div{F_1\rightarrow {50\over 12}(t+\bar t)={50\over 12}{\rm
log}|\psi^5|^2}
We also need the fact that $G_{\psi \bar \psi}$
is regular at $\psi =0$ but ${\rm exp}[K]$ goes like $|\psi|^{-2}$
near this point and that as $\psi \rightarrow \infty$,
$G_{\psi \bar \psi}^{-1}$ diverges as $|\psi |^2$ but ${\rm exp}[K]$
does not have a power law dependence on $\psi$ in this limit
\cand .  Now imposing regularity of $F_1$ at
$\psi =0$ and the requisite divergence at $\psi =\infty $
\div\ we arrive at our final result for $F_1$:
\eqn\quinf{F_1={\rm log}\left\{G_{\psi \bar \psi}^{-1}{\rm exp}\Big[{62\over
3}K\Big]
\big| \psi^{62\over 3}\big(1-\psi^5\big)^{-{1\over 6}}\big|^2\right\}}
{}From what we mentioned before, this expression should contain
in it the number of holomorphic maps from the torus to the quintic!
All we have to do is to fix $\psi$ but take the $\bar \psi
\rightarrow \infty$ limit.  Let us discuss this in full generality
for arbitrary threefold
first and then apply it to the quintic as a special case.

We want to consider the
behaviour of $F_1$ in the limit in which the `anti--holomorphic volume'
goes to infinity. The key formula is
\eqn\defK{e^{-K}= \varpi^\dagger \Omega \varpi,}
where $\varpi$ is the K\"ahler
period (\kper) in a symplectic basis.
$\varpi$ depends holomorphically on the K\"ahler moduli $z_\alpha$. In \defK\
$\Omega$ is the standard symplectic matrix.

If $t_i$ are `good coordinates' (in the sense of special geometry),
the \kper\ map takes the form
$$\varpi^t=(X_0,X_i,\partial_0 \CF,\partial_i \CF),$$
where $t_i=X_i/X_0$ and $\CF$ is
the K\"ahler prepotential. Here it is convenient to use the homogeneous
coordinates $X_I$ because
$\varpi$ takes value in some line bundle
$\CL$: $X_0$ corresponds
to a choice of trivialization of
$\CL$ (i.e. by a choice of gauge we can
set $X_0=1$ --- however
this is not necessarily
the most convenient gauge\foot{We shall always use a
holomorphic gauge, i.e. $X_0$ is assumed
to depend holomorphically on the moduli $z_\alpha$.}).
Then the usual considerations\foot{The analog formula
in the mirror picture (i.e. if the $t$'s
are regarded as complex deformations), is known as the Schmid
orbit theorems \ref\schmid{W. Schmid, Invent. Math. 22 (1973) 211.}.} give for
$t\rightarrow\infty$
$$\CF(X_0,X_i)={d_{ijk}X_iX_jX_k\over X_0}+c X_0^2+O(e^{2\pi it}),$$
where $d_{ijk}=\int \omega_i\wedge\omega_j\wedge\omega_k$
is the intersection form
for the $(1,1)$ forms and $c$ is some
constant (generated by the loop corrections).

Given the `factorized' form of eq.\defK,  taking the limit
$\bar t_i\rightarrow\infty$ while
keeping $t_j$ fixed is a well
defined procedure. More precisely,
we make $\bar t_i\rightarrow \bar s \bar t_i$ and
send $\bar s$ to infinity. In this limit
one has (up to exponentially small terms)
$$\varpi^\dagger=\bar X_0 (1, \bar s\bar t_i,
-\bar s^3 {\bar d}_{ijk}\bar t_i \bar t_j \bar t_k+2{\bar c},3\bar d_{ijk}
\bar s^2 \bar d_{ijk}\bar t_j\bar t_k).$$
Therefore
\eqn\asym{e^{-K}=\sum_{r=0}^3\bar s^r A_r,}
with
$$\eqalign{& A_3=|X_0|^2 \bar d_{ijk}\bar t_i\bar t_j\bar t_k\cr
& A_2=-3|X_0|^2 \bar d_{ijk} t_i \bar t_j \bar t_k\cr
& A_1=\bar X_0 \bar t_i \partial_i\CF\cr
& A_0=\bar X_0 \CF-2\bar c|X_0|^2.\cr}$$
Notice that in special coordinates $A_3$ and $A_2$
take a universal form (that is, they are independent of $\CF$). From \defK,
\asym\ one has
$$K=-\log X_0-\log \bar X_0 -\log[\bar s^3
d_{\bar i\bar j\bar k}\bar t_i\bar t_j\bar t_k
- 3 \bar s^2 d_{\bar i\bar j\bar k}t_i\bar t_j\bar t_k]+O(\bar s^{-2}),$$
and then
\eqn\useful{G_{\alpha\bar\beta}={1\over \bar s^2}
L_{i\bar j}(\bar t) {\partial t_i\over
\partial z_\alpha} {\partial
\bar t_j\over \partial \bar z_\beta}+O(\bar s^{-3}),}
where $L_{i\bar j}$ is a (non--singular)
{\it anti--holomorphic} matrix\foot{In fact $L_{i\bar j}$
is just the classical Zamolodchikov
metric in which we formally set $t_i=0$.}. Then
\eqn\asyG{\log\det[G_{\alpha\bar\beta}]=
\log\det[\partial t^i/\partial z^\alpha] +
{\rm anti-holomorphic}+ O(\bar s^{-1}).}
On the other hand,
\eqn\asyK{K=-\log X_0 +{\rm anti-holomorphic}+ O(\bar s^{-1}).}

The $tt^\star$ partition function $F_1$ at genus one has the form \spca\
$$F_1=\log\left[{\rm exp}{(3+h_{1,1}-{\chi
\over 12})K}\ \det[G_{\alpha \bar \beta}]^{-1}|f(z)|^2\right],$$
where $f(z)$ is a holomorphic section of $\CL^{3+h_{1,1}-{
\chi\over 12}}\otimes {\cal K}$.
As $\bar s\rightarrow\infty$ the corresponding one--point function becomes
\eqn\onepoint{\partial_\alpha F_1^{top}= \partial_\alpha\log\left[
{f(z)\over X_0^{3+h_{1,1}-{\chi \over 12}}\det(\partial t/\partial
z)}\right],} %
notice that the expression in the bracket
is just $f(z)$ normalized with
respect to the canonical section of $\CL^{
3+h_{1,1}-{\chi \over 12}}\otimes{\cal K}$, i.e.
$X_0^{3+h_{1,1}-{\chi \over 12}}\det[\partial t/\partial z]$.
Eq.\onepoint\ is the topological
one--point function at genus one. The only non--trivial ingredient in
\onepoint\ is $f(z)$.

Applying this result to the quintic case \quinf\
we get for the topological
limit
\eqn\effe{F_1^{top}=
{\rm log}\left[({\psi \over \varpi_0})^{62\over 3} (1-\psi^5)^{-{1 \over
6}}{d\psi\over
dt}\right]}
where we are using the same gauge as discussed in \cand .
{}From what we
said before \betfor\ we have
\eqn\magic{\partial_t F_1^{top}={50\over 12}- \sum_{n,r=1}^{\infty}{2nrd_r
q^{nr}\over (1-q^{nr})}
-\sum_{s=1}^{\infty}{2sn_sq^s\over 12(1-q^s)} }
where $d_r$ (resp. $n_r$) is the number of {\it primitive} elliptic (resp.
rational) curves of degree $r$. From \effe\magic\ we can read the $d_r$,
see table 1.

\vglue 12pt
\bigskip

\centerline{\vbox{\offinterlineskip
\hrule
\halign{&\vrule#&
   \strut\quad\hfil#\quad\cr
height2pt&\omit&&\omit&\cr
&Degree\hfil&&$d_r$&\cr
height2pt&\omit&\cr
\noalign{\hrule}
height2pt&\omit&&\omit&\cr
&1&&0&\cr
&2&&0&\cr
&3&&609250&\cr
&4&&3721431625&\cr
&5&&12129909700200&\cr
&6&&31147299732677250&\cr
&7&&71578406022880761750&\cr
&8&&154990541752957846986500&\cr
&9&&324064464310279585656399500&\cr
&10&&662863774391414084612496876100&\cr
height2pt&\omit&&\omit&\cr}
\hrule} }

\centerline{Table 1. \#\ of elliptic curves on the quintic 3--fold}

\bigskip

For $r\leq 3$ the $d_r$ can be computed directly, giving a check of our result.
Indeed that  $d_1=d_2=0$ follows from general facts about curves in projective
space \ref\harris{P. Griffiths and J. Harris, {\it Principles of Algebraic
Geometry}, Wiley--Interscience, New York, 1978.}. If a curve in $\BP^n$ has
degree $1$ (resp. $2$), then it is a line (resp. a plane conic) and hence
rational. In degree $3$  one has $d_3=n_2=609250$ since on a general quintic
there are as many degree 2 rational curves as degree 3 elliptic ones. In fact a
degree $3$ elliptic curve   is necessarily a plane cubic. If $C$ is such a
curve lying on the quintic, the plane containing it will meet the quintic along
a plane curve which is the union of $C$ with a conic. The higher values of
$d_i$ were not known previously.  Given the structure of \magic, it is quite
non-trivial that $d_i$ come out to be integer from the computation,
and that can be viewed as the first non-trivial check.

We have also done a similar computation on the three examples of
threefolds studied in \ref\kle{A. Klemm and S. Theisen, {\it Considerations of
One--Modulus Calabi--Yau Compactifications: Picard--Fuchs Equations, K\"ahler
Potentials and Mirror Maps}, preprints KA--THEP--03/92, TUM--TH--143--92 (April
1992).}\ and we find
the following expressions for the topological one point function
in the three cases (using their notation for labelling
the three models by $k$):
$${dF_1^{top}\over dt}=
{d\over dt}{\rm log}\left[({\psi \over \varpi_0})^{4+{204\over 12}}
{d\psi \over dt}(1-\psi ^6)^{-{1\over 6}}\right]\qquad k=6$$
$${dF_1^{top}\over dt}=
{d\over dt}{\rm log}\left[({\psi \over \varpi_0})^{4+{296\over 12}}
{d\psi \over dt}(1-\psi ^8)^{-{1\over 6}}\psi \right]\qquad k=8$$
$${dF_1^{top}\over dt}=
{d\over dt}{\rm log}\left[({\psi \over \varpi_0})^{4+{288\over 12}}
{d\psi \over dt}(1-\psi ^{10})^{-{1\over 6}}\psi\right]\qquad k=10$$
and in all three cases after the genus zero subtraction,
we get integral results for $d_r$.  The results for the first few $d_i$ are
summarized in table 2.

\vglue 12pt
\bigskip

\centerline{\vbox{\offinterlineskip
\hrule
\halign{&\vrule#&
   \strut\quad\hfil#\quad\cr
height2pt&\omit&&\omit&\cr
&Model\hfil&&$\chi$&&$\int c_2 \wedge k$&&$d_1$&&$d_2$&&$d_3$&\cr
height2pt&\omit&\cr
\noalign{\hrule}
height2pt&\omit&&\omit&\cr
&k=5&&-200&&50&&0&&0&&609250&\cr
&k=6&&-204&&42&&0&&7884&&145114704&\cr
&k=8&&-296&&44&&0&&41312&&21464350592&\cr
&k=10&&-288&&34&&280&&207680680&&161279120326560&\cr
height2pt&\omit&&\omit&\cr}
\hrule}}

\centerline{Table 2. \#\ elliptic curves on CY weighted hypersurfaces}

\bigskip

For the low values of $r$
they agree with the known results. In particular, for $k=6$ all degree $1$
curves should be `plane' rational, and thus $d_1=0$. The corresponding `plane'
meets the CY space on a curve whose other component has degree $2$; computing
its $c_1$ we see that it is a torus. Hence for $k=6$ we have $d_2=n_1=7884$.

We have generalized \ref\fut{M. Bershadsky, S. Cecotti, H. Ooguri and C. Vafa,
to appear.}\
these ideas to compute higher genus partition
functions $F_g$
of twisted $N=2$ conformal theories
coupled to gravity. The story is very similar to genus one in that $F_g$
is essentially only a function of $t$ with a simple $\bar t$ dependence
characterized by a holomorphic anomaly, relating it
to lower genus partition functions.  For example for the
special case of three-folds the anomaly is expressed by the equation
$${\bar \partial_i}F_g={\bar C_{ijk}}e^{2K}G^{j\bar j}G^{k\bar k}
[D_jD_k F_{g-1}+{1\over 2}\sum_r D_jF_r\ D_kF_{g-r}]$$
(where $F_g$ is now a section of ${\cal L}^{2g-2}$ and $D_i$
represent covariantized derivatives).  This equation can be restated as a
master equation for ${\rm exp} \sum_g (\lambda e^{2K})
^{g-1}F_g$.
Again it can be shown that
in the topological limit $\bar t \rightarrow \infty$, $F_g$
counts the number of holomorphic curves of genus $g$ in $M$.  Moreover
one can relate the above computation to some field theoretic
computation in the context of type II superstrings, thus
opening the door to the exciting possibility of
obtaining certain {\it non-perturbative} results
for superstrings using the master equation.

\vglue 16pt
We would like to thank I. Antoniadis, L. Dixon and S.T. Yau
for valuable discussions.  We would also like to thank
S. Hosono for sharing his {\it Mathematica} program for
the examples we used in this paper.  We are also grateful
to S. Katz for explaining the origin of the genus zero
contribution (see appendix B) that we had discovered experimentally.

The research of M. B., H. O. and C. V. was supported in part by
Packard fellowship and
NSF grants PHY-87-14654 and PHY-89-57162.  The research of S. C.
was supported in part by INFN. The research of H. O. was also supported
in part by Grant-in-Aid for Scientific Research on Priority Areas
231 `Infinite Analysis' from the Ministry of Education, Science and
Culture of Japan.
\vfill\eject

\appendix{A}{Contact Term Contribution}
\medskip

Let us start with the one-point function on the torus. The derivative of
$F_1$ with respect to $t_i$ brings down the integral of $\{ Q^- ,[\bar{Q}^-
,\phi_i (z) ] \}$ over the worldsheet, where $Q^\pm$ ($\bar{Q}^\pm$) are
left-moving (right-moving) $N=2$ supercharges and $\phi_i$ is
the chiral primary field of dimension zero associated to the
parameter $t_i$. If we take $Q^-$ and $\bar{Q}^-$ around the torus,
we pick up their commutator with ${\rm F_L}$ and ${\rm F_R}$ as
$$ \eqalign{
 \partial_i F_1 & = \int {d^2 \tau \over \tau_2} ~
        \Tr (-1)^{\rm F} {\rm F_L F_R}
         \left( \int d^2 z \{ Q^-,[ \bar{Q}^- , \phi_i(z) ] \}
  \right) q^{L_0} \bar{q}^{\bar{L}_0} \cr
     & = \int d^2 \tau~
         \Tr (-1)^{\rm F} Q^- \bar{Q}^- \phi_i(0) q^{L_0}
                            \bar{q}^{\bar{L}_0} \cr
     &= \int d^2 \tau~
         \Tr (-1)^{\rm F}  \left( \oint {du \over -2 \pi i} G^-(u) \right)
        \left( \oint {d\bar{u}' \over 2 \pi i} \bar{G}^-(\bar{u}')
       \right)     \phi_i(0) q^{L_0}
                            \bar{q}^{\bar{L}_0} .\cr } $$
In the last line, the supercharges are expressed as contour integrals
of the supercurrents $G^-$, $\bar{G}^-$ around the cycle of
the torus. Since the
supercurrents are (anti-) holomorphic and single-valued on the torus,
we can rewrite the above as
\eqn\onepoint{ \eqalign{ \partial_i F_1=
  \int d^2 \tau ~\Tr &\left[ (-1)^{\rm F}  \left(
   \int {d^2 u \over \pi} (\bar{\partial} \xi) G^-(u) \right)
 \left( \int {d^2 u' \over \pi} ({\partial \xi}) \bar{G}^-(\bar{u}')
        \right) \right.  \cr
  &\left. ~~~~~~~\times \phi_i(0) q^{L_0}
                            \bar{q}^{\bar{L}_0} \right] ,\cr} }
where $\xi(z,\bar{z}) = Im (z)/Im(\tau)$. This is a natural expression
for the stringy one-point function since $G^-$ and $\bar{G}^-$ are
to be viewed as reparametrization ghosts and $\bar{\partial} \xi$
is the Beltrami differential associated to the modulus
$\tau$ of the torus.

To compute the derivative of \onepoint\ with respect
to $\bar t_j$, we insert the integral of $\{ Q^+, [ \bar Q^+ ,
\phi_{\bar j}(z) ] \}$ in the trace. Again, we can take
$Q^+$ and $\bar Q^+$ around the torus, and pick up
their commutator with $G^-$ and $\bar G^-$.
$$ \eqalign{
  \partial_{\bar j} \partial_i F_1
 =
 \int d^2 \tau ~ \Tr &\left[ (-1)^{\rm F} \left( \int {d^2 u \over \pi}
     (\bar{\partial} \xi) 2 T(u) \right)
     \left( \int {d^2 u' \over \pi} (\partial \xi)
             2 \bar{T}(\bar{u}') \right) \right. \cr
  &\left. ~~~~~\times \left( \int d^2 z \phi_{\bar j} (z)
    \right) \phi_i(0)
       q^{L_0} \bar{q}^{\bar{L}_0} \right]  \cr } $$
By using the Ward identity on the torus \ref\eo{T. Eguchi
and H. Ooguri, Nucl. Phys.  B282 (1987) 308.},
we can replace the energy-momentum tensor $T$ by derivatives
with respect to $\tau$ and $z$ as
$$ \eqalign{ \partial_{\bar j} \partial_i F_1
 =& 4 \int d^2 \tau \int d^2 z
      (\partial_{\bar \tau} + \xi \partial_{\bar z}
                  + (\partial_{\bar z} \xi) )
      (\partial_\tau + \xi \partial_z + (\partial_z \xi))
         \cr
 & ~~~~~~\times \Tr (-1)^{\rm F} \phi_{\bar j}(z) \phi_i(0)
           q^{L_0} \bar{q}^{\bar{L}_0} . \cr} $$

 If the integrand were regular in $z$, we could interchange
the integral and the derivatives in the above as
$ \int d^2 z (\partial_{\bar{\tau}} + \xi \partial_{\bar z}
     + (\partial_z \xi)) \rightarrow
  \partial_{\bar \tau} \int d^2 z.$
However there is a singularity at $z=0$ due to the OPE of
$\phi_{\bar j}(z)$ and $\phi_i(0)$ as
$$ \phi_{\bar j}(z) \phi_i(0)
        \sim {G_{i \bar j} \over (2 \pi)^2 z \bar{z}, } $$
and an additional contribution arises
from the contact term
$$ \bar{\partial}[ \xi \partial ( \xi \phi_{\bar{j}}(z) ) ]
         \phi_i(0) \sim
    - {G_{i \bar j} \over 16 \pi (\tau_2)^2} \delta(z)
           + \cdots. $$
We then obtain
\eqn\contact{ \eqalign{
\partial_{\bar j} \partial_i F_1 &=
  4 \int d^2 \tau {\partial \over \partial \bar \tau}
    \int d^2 z ( \partial_\tau + \xi \partial_z + (\partial_z \xi))
     \Tr (-1)^{\rm F} \phi_{\bar j}(z) \phi_i(0) q^{L_0} \bar{q}^{\bar{L}_0}
  - \cr
 &~~~- \int {d^2 \tau \over 4 \pi (\tau_2)^2}
     ~G_{i \bar j} \Tr (-1)^{\rm F}
            q^{L_0} \bar{q}^{\bar{L}_0} .\cr}}
The second term in the right-hand side becomes
$-{1 \over 12} G_{i \bar j} \Tr (-1)^{\rm F}$ after integration over
$\tau$.

Because of the $\bar{\tau}$-derivative,
the first term in the right-hand side of \contact\ can
be expressed as an integral at $\tau_2 \rightarrow \infty$ as
$$
 \oint_{-{1 \over 2} +i \infty}^{{1 \over 2}+i \infty} d \tau
     \int d^2 z ~2i ( \partial_\tau + \xi \partial_z
         +(\partial_z \xi) )
        \Tr (-1)^{\rm F} \phi_{\bar j}(z)  \phi_i(0)
          q^{L_0} \bar{q}^{\bar{L}_0} . $$
The piece involving $\partial_\tau$ can be evaluated
using the technique of \ising\ as
$$
\eqalign{&\oint_{-{1 \over 2}+i \infty}^{{1 \over 2}+i \infty} d\tau
 \int d^2 z ~2i \partial_\tau
        \Tr (-1)^{\rm F} \phi_{\bar j}(z)  \phi_i(0)
          q^{L_0} \bar{q}^{\bar{L}_0} =  \cr
 &= \Tr (-1)^{\rm F} \left( \oint dz \phi_{\bar j}(z) \right) P
               \phi_i(0) (P-1), \cr } $$
where  $P$ is the projection operator on the ground states.
The rest is also  simplified as\foot{
The singularity at $z=0$ again generates a contact term here.
But, this time, it does not survive the $\tau_2 \rightarrow \infty$
limit.}
$$ \eqalign{
 & \oint_{-{1 \over 2} + i \infty}^{{1 \over 2} + i \infty} d\tau
  \int d^2 z ~2i(\partial_z \xi + (\partial_z \xi)) \Tr (-1)^{\rm F}
     \phi_{\bar j}(z) \phi_i(0) q^{L_0} \bar{q}^{\bar{L}_0} = \cr
 & = \Tr (-1)^{\rm F} \oint dz \phi_{\bar j}(z) P
         \phi_i(0). \cr} $$
Putting them together, the contribution from
$\tau_2 \rightarrow \infty$ is
$$ \eqalign{&\Tr (-1)^{\rm F} \left( \oint dz \phi_{\bar j}(z) \right) P
               \phi_i(0) (P - 1)
           + \Tr (-1)^{\rm F} \left( \oint dz \phi_{\bar j}(z) \right) P
               \phi_i(0) P \cr
       &= \Tr (-1)^{\rm F} \left( \oint dz \phi_{\bar j}(z) \right) P
               \phi_i(0)  P \cr
       & = \Tr (-1)^{\rm F} C_i \bar{C}_j .\cr }$$

Thus we find that the holomorphic anomaly is
expressed as
$$ \partial_{\bar j} \partial_i F_1 =
 \Tr (-1)^{\rm F} C_i \bar{C}_j - {1 \over 12} G_{i \bar j}
          \Tr (-1)^{\rm F}. $$
The second term in the right-hand side, which comes from
the contact term between $\phi_i$ and $\phi_{\bar j}$, was
missing in the previous paper \ising.

\appendix{B}{Intersection Theory over Moduli Spaces of Degenerate Instantons}
\medskip
\centerline{by Sheldon Katz\foot{
Department of Mathematics, Oklahoma State University,
Stillwater, OK 74078;\hfill \break
katz@math.okstate.edu.}}
\medskip

The computation of $n$~point functions via mirror manifolds has led to
predictions for the number of curves of certain types on Calabi-Yau
manifolds $X$.  The resulting finite numbers are independent of the complex
structure of $X$.  It is possible for $X$ to actually contain infinitely
many curves of the type in question for some complex structures and finitely
many curves for others.  This has led to the creation of an algebro-geometric
method for calculating what the number would be if it were finite, even if
the calculation took place using the ``wrong'' complex structure
\ref\eq{S. Katz, {\it Excess Intersection and Deformations}, in
preparation.}.
This number will be called the {\it contribution\/} of the family of
curves in question.  Evidence
is emerging that even if there are infinitely many curves of the type
considered for {\it any\/} complex structure on $X$, the number obtained by
algebraic geometry coincides with the number obtained by an asymptotic
expansion of the $n$~point function \ratcurves\ref\msri{S. Katz, in Essays
on Mirror Manifolds, edited by S.-T.~Yau, International Press, Hong Kong
1992.}.  In this appendix, we give
another example of this phenomenon.

Here are the main points which have arisen in the algebro-geometric
investigation.

\bigskip\noindent
1. We must consider {\it degenerate\/} instantons.  This is necessary because
the requirement of invariance of the number under variation of complex
structure forces integration over compact sets (to prevent curves from
``going off to infinity'' as the complex structure parameters of $X$ vary).
Including degenerate instantons will always compactify the space of instantons.

\medskip\noindent
2. Deformation theory tells one how to deform a holomorphic map $f:C\to X$.
The tangent space to the space of all such maps is the space of global sections
$T^0=H^0(C,f^*T_X)$, where $T_X$ is the holomorphic tangent bundle of $X$.
There is an analytic map ${\r}_f:U\to T^1$
from a neighborhood $U$ of the origin in $T^0$ to
another finite dimensional vector space $T^1$.  The vector space $T^1$ is
called the {\it obstruction space}, and ${\r}_f$ is the {\it obstruction
mapping}.  The actual space of holomorphic maps is locally described as
$\r_f^{-1}(0)$.  In this local description, the origin of $T^0$ corresponds to
the map $f$ itself.

\medskip\noindent
3. As $f$ varies, the varying $T^1$ spaces form an {\it obstruction bundle}
$\T^1$ (possibly with singularities) over the space of all maps.  As the
complex structure of $X$ is infinitesimally altered, there arises an
obstruction section \r\ of $\T^1$, whose zero locus is precisely the locus
of maps which can be deformed so as to remain holomorphic after the change
in complex structure.  From now on, $C$ will be a complex curve.
If the dimension of the zero locus of \r\ is what
one expects (in the present context, this means 3 if $C$ has genus 0,
1 if $C$ has genus 1, or 0 if $C$ has genus $g>1$), then its homology class
may be calculated via the use of Chern classes.  In this way, we can
effectively calculate the contribution of curves on $X$ by essentially doing
the computation on an infinitesimally nearby complex structure.
It may be that the obstruction sections
never yield the expected dimensions for {\it any\/} complex structure.  This
is the situation for example if the mappings are $d$~to 1 covers of a curve
in $X$, for $d>1$.  However, the Chern class may always be calculated; and
it will always have the correct dimension.  In such an instance, one
{\it assumes\/} that this Chern
class represents what the homology class ``should be'' if there indeed
were a deformation of complex structure general enough to yield finitely
many curves, and then proves under various assumptions that this is
invariant under deformations.  By analogy with excess intersection
theory \ref\fult{W. Fulton, Intersection Theory, Springer-Verlag,
Berlin Heidelberg New York 1984.},
this class is called the {\it equivalence\/} of the family of
curves in question \eq.

\medskip\noindent
4. In calculating the $n$~point function, we usually count the number of
instantons for which the images of the $n$ points lie in various hypersurfaces
determined by the $(1,1)$ forms under consideration.  Here, there is an
analogous condition for all (possibly degenerate) instantons.  The condition
is reinterpreted as a cohomology class, which can be intersected with the
result of the Chern class calculation in the last step to give a number, the
contribution of the family.

\medskip
We want to apply these ideas to calculate the equivalence of the family
of degenerate instantons corresponding to maps from an elliptic curve to
a rational curve.

\bigskip
Now, for specifics.  Motivated by the computation in \ratcurves, one
identifies $f:C\to X$ with its graph $\G_f\subset C\times X$.  It is well
known in algebraic geometry how to compactify the space of all subvarieties
of a given algebraic variety (here $C\times X$): one uses the {\it Hilbert
scheme} $H=\hbox{Hilb}(C\times X$) which parametrizes {\it all\/} subvarieties
of
$C\times X$ \ref\groth{A. Grothedieck, Fondements de la G{\'e}ometrie
Alg{\'e}brique, S{\'e}minaire Bourbaki, Secr{\'e}tariat Math., Paris
1962.}.
Here, we restrict our attention to the connected component
of $H$ which contains $\{\G_f\}$.  Assume Im$(f)=D\subset X$ is a curve
which is rigid in $X$.  Then all deformations of $f$ will continue to map
inside $D$, and we may as well consider Hilb$(C\times D)$.  Let $\p_C$
and $\p_D$ be the projection mappings of $C\times D$, and let $c\in C$ and
$d\in D$ be general points.  Then if $f:C\to D$ is a degree $k$ covering,
we have
$\p_C^{-1}(c)\cdot\G_f=1$ and $\p_D^{-1}(d)\cdot\G_f=k$.  Also, $\G_f$ has
the same genus as $C$.  So we compactify by including all subvarieties \G\
of $C\times D$ such that $\G\cdot\p_C^{-1}(c)=1,\ \G\cdot\p_D^{-1}(d)=k$,
and $g(\G)=g(C)$.

Now, let $C$ be an elliptic curve, and let $D\subset X$ be a rigid
rational curve, fixed for the remainder of this appendix.
There are no degree 1 maps from $C\to D$.  But there are still singular
subvarieties of $C\times D$ as above that must be considered as degenerate
instantons.  Let $p\in C$ and $q\in D$ denote arbitrary points.  Let
$\G_{pq}=(\{p\}\times D)\cup (C\times \{q\})\subset C\times D$.  Then
$\G_{pq}$ has genus 1, and $\G_{pq}\cdot\p_C^{-1}(c)=1,\
\G_{pq}\cdot\p_D^{-1}(d)=1$.  It is easy to see that these are the only
possible degenerate instantons.

Before we calculate the obstruction bundle, we must identify the
moduli space of degenerate instantons globally.  First of all, we
must not only specify
an elliptic curve $C$ and points $c\in C,\ d\in D$, but also a marked point
$p_1\in C$, since we need to consider pointed elliptic curves for the
1~point function.  It is well known that there is no consistent way to
describe what one might think of as $\{(E,p)|p\in E\}$.  This is because of
the presence of automorphisms.  Given any elliptic curve $E$, the group of
automorphisms acts transitively; hence modulo automorphisms, any point of
$E$ is the same as any other point.  However, a point cannot be selected from
$E$ in a continuous fashion as $E$ varies due to the presence
of extra automorphisms that certain elliptic curves possess.  This situation
is usually remedied by introducing a ``level $k$'' structure on $E$ for some
$k\ge 3$. Concretely for $k=3$, this means we consider the hypersurface \C\ in
$\P1\times\P2$ defined by the equation
\eqn\level{s(x^3+y^3+z^3)+txyz=0.}
Here $(s:t)$ are homogeneous coordinates for \P1, and $(x:y:z)$ are
homogeneous coordinates for \P2.  Via the projection map $\p:\C\to\P1$, this
is thought of as a family of plane
cubic curves parametrized by $(s,t)\in\P1$.  Each curve in the family \C\
contains the marked point $p_1=(1,-1,0)$; so we have succeeded in giving a
family
of pointed elliptic curves.  The problem is that we have described all
elliptic curves multiply; consideration of the $j$~invariant of \level\
shows that each
elliptic curve occurs 12 times in this family (including multiplicity).
So the number which results from using \C\ for computational purposes must
be divided by 12 at the end.

We are now ready to compute the moduli space of degenerate instantons.
We need to specify a pointed elliptic curve $(E,p_1)$, a point $p\in E$, and a
point $q\in D$.  The data $(E,p_1,p)$ is just the specification of a point
of \C\ (modulo the 12 to 1 identifications given by the $j$~invariant, which
will not be mentioned again until the last step).  So the moduli space is in
this instance just $\C\times D$.  Note that this space is nonsingular of
dimension 3.  A 1~dimensional class (the equivalence)
must be obtained before imposing the
condition on $p_1$; hence we will need to calculate a second Chern class.

The cohomology ring of $\C\times D$ may be easily worked
out.  By the
K\"unneth formula, the result is just the tensor product of the
cohomology
rings of \C\ and $D$.  We write $h$ for the positive integral generator
of $H^2(D)$.  \C\ is well known to be isomorphic to the blow up of \P2\ at the
base locus of the pencil of elliptic curves defining \C, i.e.\ the nine points
$x^3+y^3+z^3=xyz=0$.  So $H^2(\C)$ is generated by classes $H,E_1,\ldots,E_9$,
where $H$ is the pullback of the hyperplane class of \P2\ to \C\ and the
$E_i$ are the classes of the
exceptional divisors.  We order the $E_i$ so that $E_1$ corresponds to the
marked point $p_1$.
Note that the pullback via \p\ of the hyperplane class $a$
of \P1 is just $3H-\sum_{i=1}^9E_i$.

Putting the preceding together, we see that the instantons which we consider
here are identified with certain subvarieties of $C\times D$, as $C$ varies
over all pointed elliptic curves.  This is what is meant by the
{\it relative Hilbert scheme\/} Hilb$(\C\times D/\P1)$ \groth.
We specialize now to $k=1$.  We have seen that the degenerate
instantons are of the form $\G_{pq}=(\{p\}\times D)\cup (C\times\{q\})$, where
$p\in C$ and $q\in D$  and are
parametrized by $\C\times D$.  We want to describe $\G_{pq}$ globally.

Consider the space $\CC\times D\times D$.
There are diagonals $\D_{\cal C}\subset \CC$ and $\D_D\subset
D\times D$. Let $\G=(\D_{\cal C}\times D\times D)\cup(\CC\times\D_D)$.
Let $\p_1,\ \p_2$ be the projections onto the respective \C\ factors, and let
$\r_1,\ \r_2$ be the projections onto the respective $D$ factors.  Then if
$C$ is a curve of the family \C, and if $p\in\C$ and $q\in D$, we have
$\G\cap(\p_1\times\r_1)^{-1}\{(p,q)\}=\G_{pq}$.  So \G\ is the
``universal degenerate instanton'' parametrized by $\C\times D$.

Note that while \C\ is smooth, the fibers of $\C/\P1$ can be singular.  In
fact, we easily compute from \level\ that
there are 3 singular points over $s=0$, and one singular point over
each of the three points with $t/s=-3e^{2\pi im/3}$, for $m=0,1,2$.  Thus
$\CC$ itself has singularities---there are 9 singular points
over $s=0$, and one singular point over each point $t/s=-3e^{2\pi im/3}$, for
$m=0,1,2$ \ref\sch{C. Schoen, Math.\ Z. 197 (1988) 177.}.  Although the
presence
of singularities can often complicate the method of \eq, we will see presently
that they cause no problem in this case.

In general terms, the obstruction space $T^1$ at $\G_{pq}$ is computed in
terms of the normal sheaf of $\G_{pq}$ in $C\times X$ \groth\msri.  Note that
for the graph $\G_f$ of an actual holomorphic mapping $f:C\to X$, the normal
sheaf of
$\G_f$ in $C\times X$ is just $f^*(T_X)$.  This should be compared with
\ratcurves.  The normal sheaf of a degenerate instanton may differ
from the pulled back tangent bundle considered in \ratcurves; however, when
the present obstruction analysis is applied to the calculation of the
contribution to the 3~point function as in \ratcurves, the identical result
is obtained.

The normal sheaf roughly speaking has two parts: the normal sheaf of
$\G_{pq}$ in $C\times D$, and the pullback to $\G_{pq}$ via $\r_2$ of the
normal bundle of $D$ in $X$.  Note that since we have assumed that $D$ is
rigid, we know that the normal bundle of $D$ in $X$ is $\O_D(-h)\oplus
\O_D(-h)$.

There are no relevant obstructions associated to the normal sheaf of
$\G_{pq}$ in $C\times D$.  We can describe the deformations of $\G_{pq}$
by deforming $p$ in $C$ and $q$ in $D$.  There are no obstructions to
deforming $q$, and there are no obstructions to deforming $p$ unless $p$
is a singular point of a singular fiber of \C.  But even in this case,
the only obstructions are the obstructed tangent directions at the singular
point; but these are of no concern to us.

So the only relevant
obstructions come from the cohomology sheaf of the pullback
of the sheaf $\O_D(-h)\oplus\O_D(-h)$.  We may restrict attention to one of
these
factors; and we must calculate the sheaf
$R^1((\p_1\times\r_1)\mid_{\Gamma})_*(\r_2^*\O_D(-h))$
on $\C\times D$.  We check first that this is a rank 1 bundle on $\C\times D$,
and then that it is equal to $\O_{{\cal C\times D}}(-a-h)$, where we have
simplified notation by supressing the pullbacks of $a$ and $h$.

We first note that $\dim H^1(\G_{pq},\r_2^*\O_D(-h))=1$ for all
$p$ and $q$, even for $p$ a singular point of a singular fiber.  In fact,
the restriction of $\r_2^*\O_D(-h)$ to $C\times\{q\}$ is trivial, so that
$\dim H^1(C\times\{q\},\r_2^*\O_D(-h))=1$. We also see that the restriction
map is an isomorphism by considering the exact cohomology sequence associated
to the short exact sequence
$$0\to\O_D(-2h)\to\r_2^*\O_D(-h)\to\O_C\to 0,$$
where the right hand map is restriction.  The sheaf on the left arises as the
subsheaf of $\O_D(-h)$ consisting of sections vanishing at $\{q\}$.  We
have simplified notation by using $C$ and $D$ to stand for $C\times\{q\}$
and $\{p\}\times D$, respectively.

This shows that $R^1((\p_1\times\r_1)\mid_{\Gamma})_*
(\r_2^*\O_D(-h))$ is a line
bundle on $\C\times D$ \ref\hart{R. Hartshorne, Algebraic Geometry,
Springer-Verlag, New York Berlin Heidelberg 1977.} (whose fiber over
$\{(p,q)\}$ is $H^1(\G_{pq},\r_2^*\O_D(-h))$).

The above argument also shows that
\eqn\image{R^1((\p_1\times\r_1)\mid_{\Gamma})_*\r_2^*\O_D(-h)
\simeq
R^1((\p_1\times\r_1)\mid_{{\cal C}\times_{{\bf P}^1}{\cal C}\times\Delta_D})_*
\r_2^*\O_D(-h),}
since we have just shown that this isomorphism holds fiber
by fiber.

Since we have restricted $D\times D$ to $\D_D$, we may replace $\r_2$ by
$\r_1$ in \image\ and compute instead
\eqn\project{R^1((\p_1\times\r_1)\mid_{{\cal C}\times_{{\bf P}^1}{\cal C}
\times\Delta_D})_*\r_1^*\O_D(-h)\simeq
R^1{\p_1}_*(\O_{{\cal C}\times_{{\bf P}^1}{\cal C}\times\Delta_D})\otimes
\p_D^*\O_D(-h)}
by the projection formula.  Since higher direct images commute with base
extension in this instance \hart, the desired result will follow from
\image\ and \project\ once we show
that $R^1\p_*\O_{{\cal C}}\simeq\O_{{\bf P}^1}(-a)$.  This follows for
example by first noting that $\dim H^1(\O_C)=1$ for all fibers $C$ of \p,
showing
that $R^1\p_*\O_{{\cal C}}$ is a line bundle on \P1.  To identify the
line bundle, one can apply the Grothendieck-Riemann-Roch formula \fult.
The straightforward calculation has been checked
using ``schubert'' \ref\schub{S. Katz and S. A. Str\o mme, ``schubert'':
a Maple package for intersection theory, Available by anonymous ftp
from ftp.math.okstate.edu or linus.mi.uib.no, cd pub/schubert.}.

So the obstruction bundle $\T^1$ is just $\O(-a-h)\oplus\O(-a-h)$.  Its second
Chern class is $(-a-h)^2=2ah$.  This class is the equivalence of the family
of instantons.


We next turn to the ``1~point condition''.  That is, we choose a hyperplane
section $P$ of $X$ which is a representative of the $(1,1)$ class that we are
interested in\foot{Notation in this appendix differs slightly from notation
in the paper. A single $(1,1)$ class is considered here for notational
simplicity, rather than effectively considering all classes simultaneously
by grouping curves by their homology classes rather than their degrees.},
and look at the set of
all \G\ which contain a point of $\{p_1\}\times P$. This is a restatement in
terms of graphs of the usual condition $f(p_1)\in P$. If $D\subset X$ has
degree $s$ (relative to our $(1,1)$ class), then $D$ intersects $P$ in $s$
points.  Let $q$ be one of these
points.  Then we just have to
look at the set of all \G\ which contain the point $(p_1,q)$, and multiply this
class (independent of $q$) by $s$.  Before multiplying, this subvariety is
clearly just $(E_1\times D)\cup
(\C\times\{q\})$.  Its class is just $E_1+h$.

We finally impose the 1~point condition on the equivalence.
Recalling that $\p^*(a)=
3H-\sum_{i=1}^9E_i$, we get the following contribution from the family of
degree 1 degenerate maps from the fibers of \C\ to $D$.
$$\eqalign{2ah(E_1+h)&=2(3H-\sum_{i=1}^9E_i)h(E_1+h)\cr
                     &=-2E_1^2h\cr
                     &=2\cr}$$

Putting this all together, we get ${2\over 12}sn_s$ as the contribution of
the family of degenerate maps with covering degree 1
from elliptic curves to all degree $s$ rational curves.
This gives a contribution of ${2\over 12}sn_se^{-ts}$ to the
1~point function.  Here $n_s$ is the number of rational curves of degree $s$
in $X$, all assumed rigid.

The calculation of the contribution of degree $k$ covers of a rational curve
is more complicated, since the space of degree $k$ instantons is singular.
But the above analysis makes it clear that the result will be of
the form ${a_k\over 12}sn_s$ for some integers $a_k$, independent of the
choice of Calabi-Yau manifold.  This yields the formula
$${1\over 12}\sum_ksn_sa_ke^{-tks}$$
for the contribution to the 1~point function of all (possibly degenerate) maps
from elliptic curves to rational curves.  We have shown that $a_1=2$.
This analysis is consistent with \rational ; in the language of this
paragraph, \rational\ asserts that $a_k=2$ for all $k$, but this has not
yet been checked by algebraic geometry---the obstruction analysis needed in
the general case is more delicate.

We would like to thank D.R.~Morrison for numerous conversations relating to
degenerate instantons.  We would also like to thank A.~Yukie for a helpful
discussion about the universal level~3 elliptic curve.

\listrefs
\end